\documentclass[doublecol]{epl2}
\title{Link Prediction Based on Local Random Walk}
\author{Weiping Liu \and Linyuan L\"{u}\footnote{Corresponding author: linyuan.lue@unifr.ch}}
\shortauthor{Weiping Liu and Linyuan L\"{u}}

\institute{
  Department of Physics, University of Fribourg, Chemin du
Mus\'{e}e 3, CH-1700 Fribourg, Switzerland} 

\abstract{The problem of missing link prediction in complex networks
has attracted much attention recently. Two difficulties in link
prediction are the sparsity and huge size of the target networks.
Therefore, the design of an efficient and effective method is of
both theoretical interests and practical significance. In this
Letter, we proposed a method based on local random walk, which can
give competitively good prediction or even better prediction than
other random-walk-based methods while has a lower computational
complexity.}
\begin{document}

\maketitle

\section{Introduction}
Recently, the problem of missing link prediction in complex network
has attracted much attention \cite{Clauset2008, Kleinberg2007,
Getoor2005}. Link prediction aims at estimating the likelihood of
the existence of a link between two nodes. For some networks,
especially biological networks such as protein-protein interaction
networks, metabolic networks and food webs, the discovery of links
(i.e., interactions) is costly in the laboratory or the field, and
thus the current knowledge of those networks is substantially
incomplete \cite{Martinez1999,Sprinzak2003}. Instead of blindly
checking all the possible interactions, predicting based on the
observed interactions and focusing on those links most likely to
exist can sharply reduce the experimental costs if the predictions
are accurate enough \cite{Clauset2008}. For some others like
web-based friendship networks, very likely but not yet existent
links can be suggested to users as recommendations of promising
friendships, which can help users in finding new friends and thus
enhance their loyalties to the web sites. In addition, the link
prediction algorithms can be applied to solve the classification
problem in partially labeled networks \cite{Gallagher2008}, such as
to distinguish the research areas of scientific publications.

Commonly, two nodes are more likely to be connected if they are more
similar, where a latent assumption is that the link itself indicates
a similarity between the two endpoints and this similarity can be
transferred through the links. In this case, the similarity indices
are used to quantify the structural equivalence (see, for example
the Leicht-Holme-Newman index \cite{Leicht2006} and the transferring
similarity \cite{Sun2009}). However, in some networks the two
endpoints of one link are not essentially similar, such as the
sexual network \cite{Liljeros2001} and the word networks. In these
cases, we can use the regular equivalence (see Ref.
\cite{White1983}, for mathematical definition of regular
equivalence), which indicates that two nodes are said to be similar
if they have connected to the similar nodes. How to predict missing
links in such kind of networks is still an open problem to us. Our
study focuses on structure equivalence.

Node similarity can be defined by the essential attributes of nodes.
For example, if two persons have same age, sex and job, we can say
that they are similar. Another group of similarity is based only on
the network structure. An introduction and comparison of some
similarity indices is presented in Ref. \cite{Kleinberg2007}, where
the \emph{Common Neighbours} \cite{Lorrain1971}, \emph{Jaccard
coefficient} \cite{Jaccard1901}, \emph{Adamic-Adar Index}
\cite{Adamic2003} and \emph{Preferential Attachment}
\cite{Barabasi1999} are the node-dependent indices that require only
the information about node degree and the nearest neighborhood,
while the \emph{Katz Index} \cite{katz1953}, \emph{Hitting Time}
\cite{Gobel1974}, \emph{Commute Time} \cite{Fouss2007}, \emph{Rooted
PageRank} \cite{Brin1998}, \emph{SimRank} \cite{Widom2002} and
\emph{Blondel Index} \cite{Blondel2004} belong to the path-dependent
indices that ask for global knowledge of the network topology. In
Ref. \cite{Linyuan2009}, Zhou \emph{et al.} proposed two new local
indices, \emph{Resource Allocation index} and \emph{Local Path
index}. Empirical results show that these two indices perform very
well among all known local indices. In particular, the local path
index, asking for a little bit more information than common
neighbours, provides competitively accurate prediction compared with
the global index \cite{Linyuan20092}. L\"{u} and Zhou
\cite{Linyuan_weak} studied the link prediction problem in weighted
networks, and found that the weak links may play a more important
role than strong links. Besides, Clauset \emph{et al.}
\cite{Clauset2008} proposed an algorithm based on the hierarchical
network structure, which gives good predictions for the networks
with hierarchical structures, such as grassland species food web and
terrorist association network. In real application, similarity
indices only exploiting local information are more efficient than
those based on global information, for their lower computational
complexity. However, due to the insufficient information, local
indices may be less effective for their low prediction accuracy. To
design an efficient and effective algorithm is a main challenge in
link prediction.

In this Letter, we define the node similarity based on local random
walk, which has lower computational complexity compared with other
random-walk-based similarity indices, such as average commute time
(ACT) and random walk with restart (RWR). We compare our method with
five representative similarity indices, including three local ones
(common neighbours, resource allocation and local path indices) and
two global ones (ACT and RWR), as well as the hierarchical structure
method. Empirical results on five real networks show that our method
performs best.

\begin{table*}
\caption{The basic topological features of the giant components of
the five example networks. $N$ and $|E|$ are the total numbers of
nodes and links, respectively. $\langle{k}\rangle$ is the average
degree of the network. $\langle{d\rangle}$ is the average shortest
distance between node pairs. $C$ and $r$ are clustering coefficient
\cite{Watts1998} and assortative coefficient \cite{Newman2002},
respectively. $H$ is the degree heterogeneity, defined as
$H=\frac{\langle k^2\rangle}{\langle k\rangle^2}$.}
\begin{center}
\begin{tabular} {cccccccc}
  \hline \hline
   Networks     & $N$  &  $|E|$  &  $\langle k\rangle$ & $\langle d\rangle$ & $C$ & $r$ & $H$ \\
   \hline
   USAir & 332 & 2126 & 12.807 & 2.46 & 0.749  & -0.208 & 3.464 \\
   NetScience & 379 & 941 & 4.823 & 4.93 & 0.798 & -0.082 & 1.663 \\
   Power & 4941 & 6594 & 2.669 & 15.87 & 0.107  & 0.003 & 1.450 \\
   Yeast & 2375 & 11693 & 9.847 & 4.59 & 0.388 & 0.454 & 3.476 \\
   C.elegans & 297 & 2148 & 14.456 & 2.46 & 0.308  & -0.163 & 1.801 \\
   \hline \hline
    \end{tabular}
\end{center}
\end{table*}

\section{Similarity Based on Local Random Walk}
Consider an undirected simple network $G(V,E)$, where $V$ is the set
of nodes and $E$ is the set of links. Multiple links and
self-connections are not allowed. For each pair of nodes, $x,y\in
V$, we assign a score, $s_{xy}$. In this Letter, we adopt the
simplest framework, that is, to directly set the similarity as the
score. All the nonexistent links are sorted in descending order
according to their scores, and the links at the top are most likely
to exist.

Random walk is a Markov chain describing the sequence of nodes
visited by a random walker \cite{Kemeny1976,Norris1997}. This
process can be described by the transition probability matrix, $P$,
with $P_{xy}=a_{xy}/k_{x}$ presenting the probability that a random
walker staying at node $x$ will walk to $y$ in the next step, where
$a_{xy}$ equals 1 if node $x$ and node $y$ are connected, 0
otherwise, and $k_{x}$ denotes the degree of node $x$. Given a
random walker starting from node $x$, denoting by $\pi_{xy}(t)$ the
probability that this walker locates at node $y$ after $t$ steps, we
have
\begin{equation}
\vec{\pi_x}(t)=P^{T}\vec{\pi_x}(t-1),
\end{equation}
where $\vec{\pi_x}(0)$ is an $N\times{1}$ vector with the $x^{th}$
element equal to 1 and others all equal to 0, and $T$ is the matrix
transpose. The initial resource is usually assigned according to the
importance of nodes \cite{Zhou2008}. Here, we simply set the initial
resource of node $x$ proportional to its degree $k_x$. Then, after
normalization the similarity between node $x$ and node $y$ is
\begin{equation}
s^{LRW}_{xy}(t)=\frac{k_x}{2|E|}\cdot\pi_{xy}(t)+\frac{k_y}{2|E|}\cdot\pi_{yx}(t),
\label{LRW}
\end{equation}
where $|E|$ is the number of links in the network. It is obvious
that $s_{xy}=s_{yx}$. Note that, here we only focus on the few-step
random walk not the stationary state which can be characterized by
the eigenvector centrality \cite{Bonacich1987,Noh2004}. In the
stationary state, we have $\pi_{xy}=\frac{k_x}{2|E|}$, and thus
according to Eq.~\ref{LRW}, $s_{xy}=\frac{k_x\cdot{k_y}}{2|E|^2}$,
which is equivalent to the preferential attachment index (i.e.,
$k_x\cdot{k_y}$) that has been discussed in Ref. \cite{Linyuan2009}.

One difficulty with all random-walk-based similarity measures is
their sensitive dependence to parts of the network far away from
target nodes \cite{Kleinberg2007}. For example, in a random walk
from $x$ to $y$, the walker has a certain probability to go too far
away from both $x$ and $y$ although they may be close to each other.
This may lead to a low prediction accuracy since in most real
networks nodes tend to connect with the ones nearby rather than far
away. This feature relates to the high clustering or locality of
networks. A possible way to counteract this dependence is to
continuously release the walkers at the starting point, resulting in
a higher similarity between the target node and the nodes nearby. By
superposing the contribution of each walker (walkers move
independently), we obtain the similarity index:
\begin{equation}
s_{xy}^{SRW}(t)=\sum_{l=1}^{t}s_{xy}^{LRW}(l),
\end{equation}
where SRW is the abbreviation for superposed random walk.

\section{Metrics}
To test the algorithm's accuracy, the observed links, $E$, are
randomly divided into two parts: the training set, $E^T$, and the
probe set, $E^P$. Clearly, $E=E^T\cup E^P$ and $E^T\cap E^P={\o}$.
We use two standard metrics, AUC\footnote{Actually, AUC is formally
equivalent to the \emph{Wilcoxon rank-sum test} \cite{Wilcoxon1945}
and \emph{Mann-Whitney $U$ statistical test} \cite{Mann1947}. It is
a non-parametric test for assessing whether two independent samples
of observations come from the same distribution. Notice that, a
latent assumption in AUC metric is the independence of the existence
of each link, which may be not the case in the real world.}
\cite{Hanely1982} and precision \cite{Herlocker2004}, to quantify
the accuracy of prediction algorithms. The former evaluates the
overall ranking resulted from the algorithm, while the later focuses
on the top-\emph{L} candidates. In the present case, AUC can be
interpreted as the probability that a randomly chosen missing link
(a link in $E^P$) is given a higher score than a randomly chosen
nonexistent link (a link in $U\setminus E$, where $U$ denotes the
universal set). In the implementation, among $n$ independent
comparisons, if there are $n'$ times the missing link having a
higher score and $n''$ times they are of the same score, we have
\begin{equation}
AUC=\frac{n'+0.5n''}{n}.
\end{equation}
If all the scores are generated from an independent and identical
distribution, the AUC should be about 0.5. Therefore, the degree to
which the AUC exceeds 0.5 indicates how much better the algorithm
performs than pure chance. Precision is defined as the ratio of
relevant items to the number of selected items. In our case, to
calculate precision we firstly need to rank all the nonexistent
links in decreasing order according to their score. Then we focus on
the top-$L$ (here $L=100$) links. If there are $l$ links
successfully predicted (i.e., in the probe set), then
\begin{equation}
Precision=\frac{l}{L}.
\end{equation}
Clearly, higher value of precision means higher prediction accuracy.
%Readers are encouraged to see the Refs.
%\cite{Geisser1993,Herlocker2004} for more information about how to
%evaluate the accuracy of prediction.

\section{Data}
We consider five representative networks drawn from disparate
fields: (i) USAir: The network of the US air transportation system,
which contains 332 airports and 2126 airlines. (ii) NetScience: A
network of coauthorships between scientists who are themselves
publishing on the topic of network science \cite{Newman2006}. This
network contains 1589 scientists, 128 of which are isolated. In
fact, it consists 268 components, and the size of the giant
component is only 379. (iii) Power Grid: An electrical power grid of
the western US \cite{Watts1998}, with nodes representing generators,
transformers and substations, and edges corresponding to the high
voltage transmission lines between them. (iv) Yeast: A
protein-protein interaction network of yeast containing 2617
proteins and 11855 interactions \cite{Mering2002}. Although this
network is not well connected (it contains 92 components), most of
the nodes belong to the giant component, whose size is 2375. (v)
C.elegans: The neural network of the nematode worm C.elegans, in
which an edge joins two neurons if they are connected by either a
synapse or a gap junction \cite{Watts1998}. In this Letter, we only
consider the giant component, because the similarity indices based
on local random walk, as well as those well-known indices (except
the preferential attachment index) reported in Refs.
\cite{Kleinberg2007,Linyuan2009}, will give zero score to a pair of
nodes located in two disconnected components. This implies that if a
network is unconnected, we actually predict the links in each
component separately, and any probe link connecting two components
can not be predicted. Therefore we need to make sure that the
training set represents a connected network. Actually, each time
before moving a link to the probe set, we first check if this
removal will make the training network disconnected. Table 1
summarizes the basic topological features of the giant components of
those networks.

\begin{table*}
\caption{Comparison of algorithms' accuracy quantified by AUC and
Precision. For each network, the training set contains 90\% of the
known links. Each number is obtained by averaging over 1000
implementations with independently random divisions of training set
and probe set. We set the parameters $\varepsilon=10^{-3}$ in LP and
$c=0.9$ in RWR.  The numbers inside the brackets denote the optimal
step of LRW and SRW indices. For example, 0.9723(2) means the
optimal AUC is obtained at the second step of LRW. The highest
accuracy in each line is emphasized by black. For HSM we generate
5000 samples of dendrograms for each implementation.}
\begin{center}
\small
\begin{tabular}{|c|cccccc|cc|}
  \hline \hline
   \textbf{AUC}    &CN &RA &LP  &ACT &RWR &HSM &LRW &SRW  \\
   \hline\hline
   USAir   &0.9542&0.9723  &0.9524   &0.9012  &0.9765   & 0.9038 &0.9723(2)  &\textbf{0.9782}(3) \\
   \hline
   NetScience  &0.9784  &0.9825   &0.9855  &0.9338   &\textbf{0.9928}  &0.9295  &0.9893(4) &0.9917(3)\\
   \hline
   Power  &0.6257  & 0.6258   & 0.6974 &0.8948   &0.7599  &0.5025  & 0.9532(16)& \textbf{0.9631}(16) \\
   \hline
   Yeast  &0.9151  &0.9163   &0.9700  &0.8997   &0.9782  & 0.6720 &0.9744(7)&\textbf{0.9801}(8)\\
   \hline
   C.elegans  &0.8492  &0.8705   &0.8672  &0.7470   &0.8888  & 0.8082 &0.8986(3)&\textbf{0.9062}(3)\\
   \hline \hline
%& & & & & & & & & \\
%  \hline \hline
   \textbf{Precision}    &CN &RA &LP  &ACT &RWR &HSM &LRW &SRW \\
   \hline\hline
   USAir   &0.5907  &0.6350   &0.6078  &0.4887   &0.6519  &0.2764  &0.6435(3)&\textbf{0.6724}(3)\\
   \hline
   NetScience  & 0.2618 &0.5442   &0.3007  &0.1911   &\textbf{0.5485}  &0.2502  &0.5442(2)&0.5442(2)\\
   \hline
   Power  &0.1121  &0.0806   &\textbf{0.1284}  & 0.0813  & 0.0863 &0.0040  &0.0806(2)&0.1140(3)\\
   \hline
   Yeast  &0.6707  &0.4949   &0.6823  &0.5680   &0.5217  &0.8408  &\textbf{0.8591}(3)&0.7268(9) \\
   \hline
   C.elegans &0.1222  &0.1266   &0.1391  &0.0654   &0.1305  & 0.0763 &0.1399(3) &\textbf{0.1407}(3) \\
   \hline \hline
\end{tabular}
\end{center}
\label{compare}
\end{table*}

\begin{figure*}
\begin{center}
\includegraphics[width=8cm]{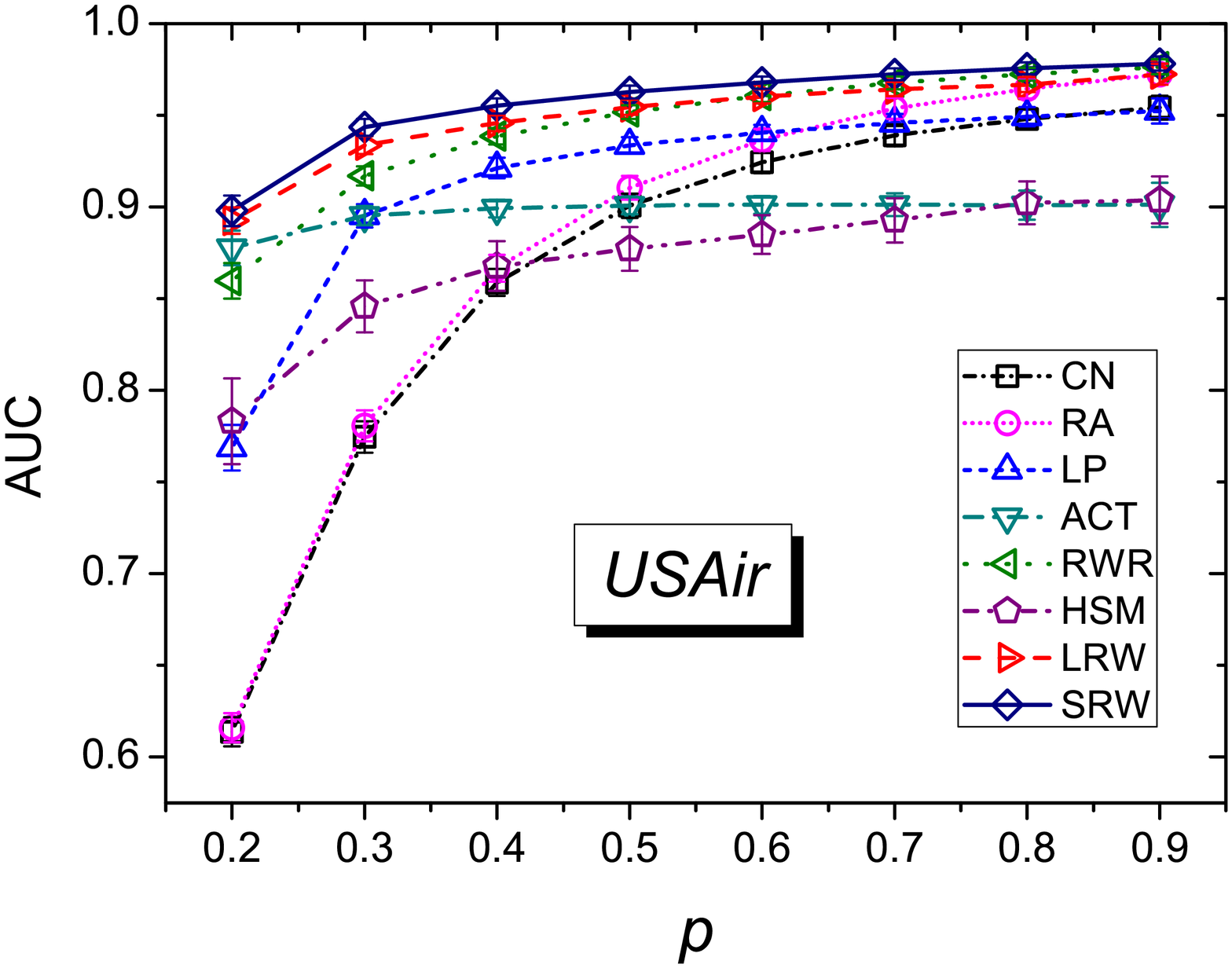}
\includegraphics[width=8cm]{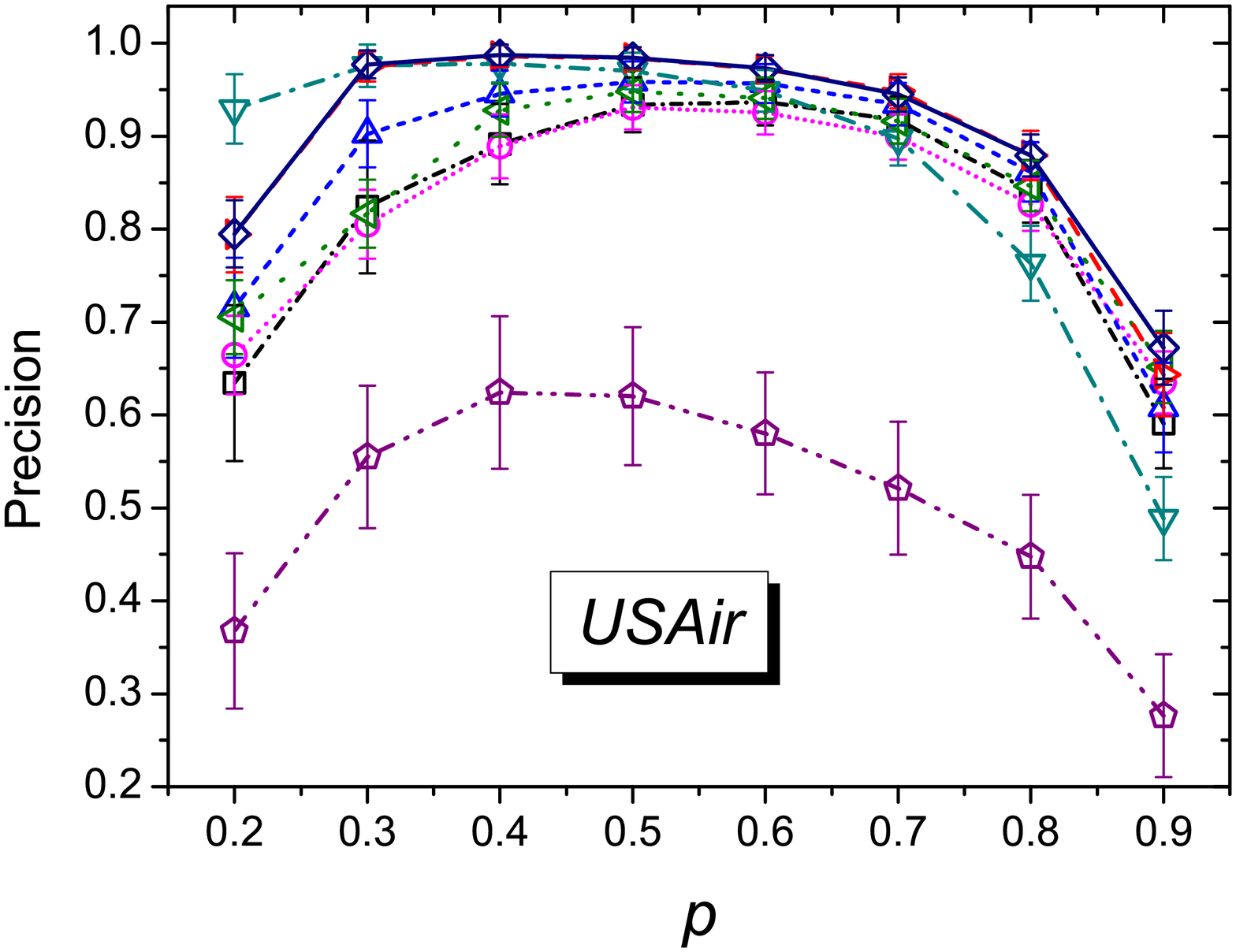}
\includegraphics[width=8cm]{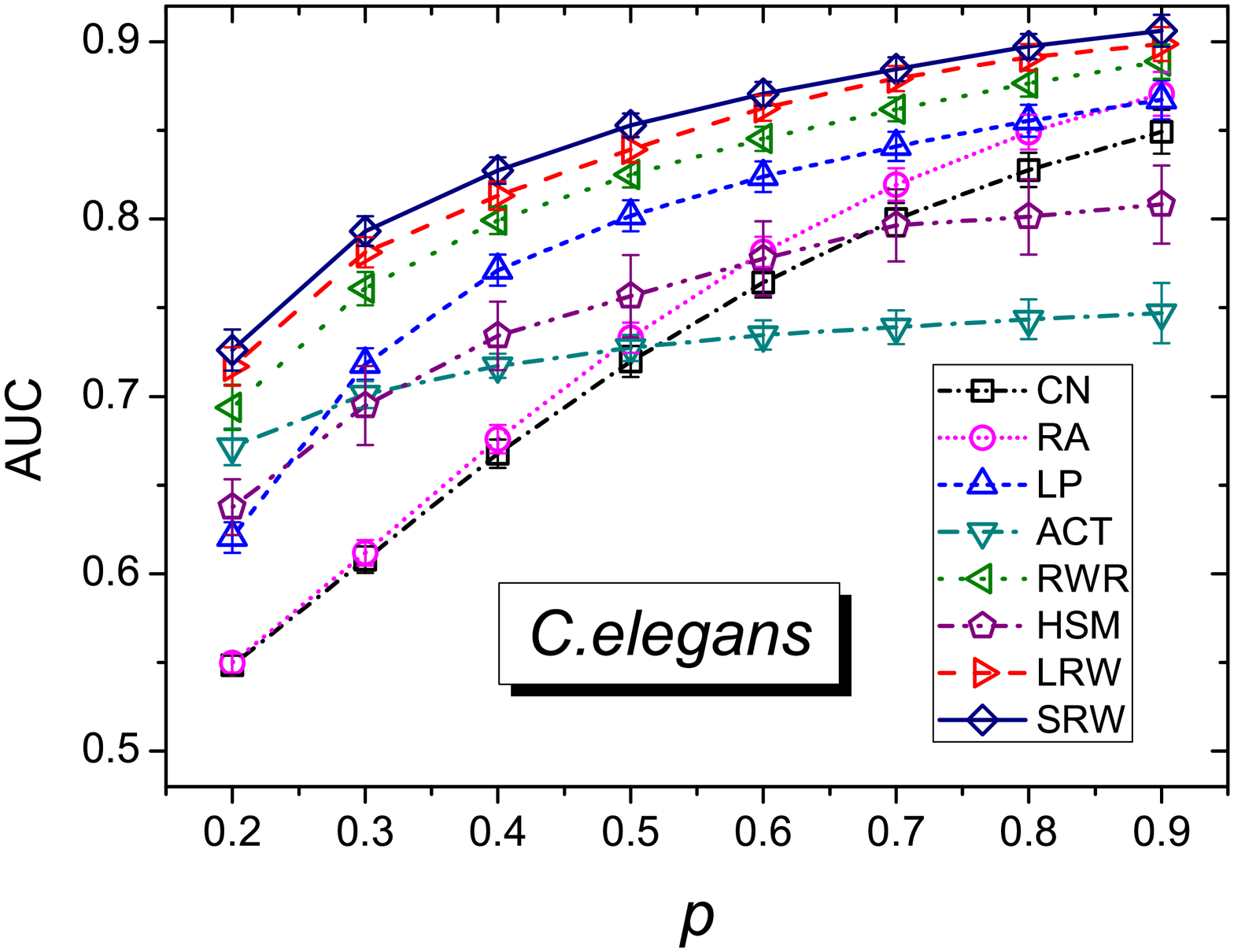}
\includegraphics[width=8cm]{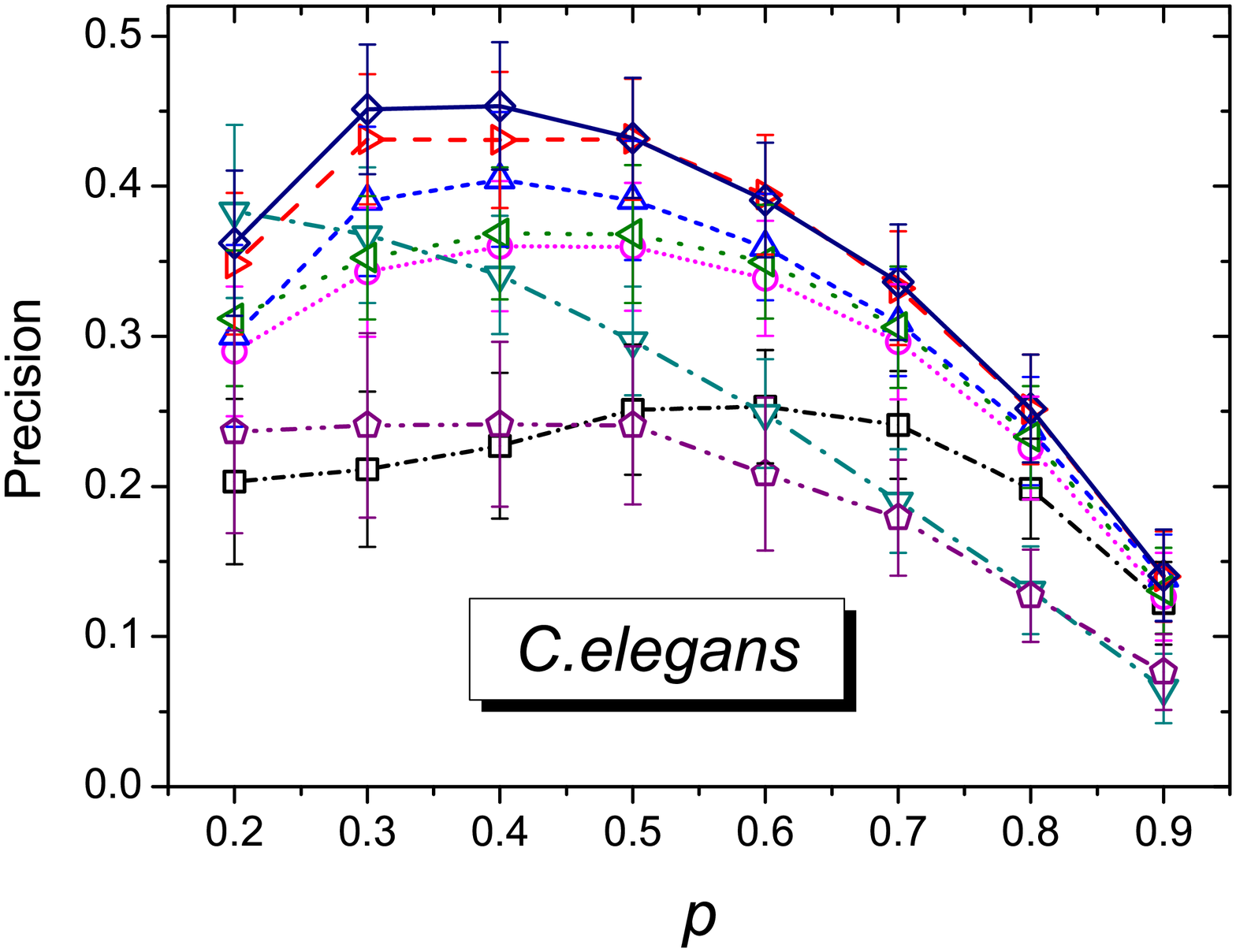}
\caption{(Color online) Dependence of AUC and Precision on the size
of training set in USAir and C.elegans. Each number is obtained by
averaging over 1000 implementations with independently random
divisions of the training set and probe set. For HSM we generate
5000 samples of dendrograms for each implementation.}\label{train}
\end{center}
\end{figure*}

\section{Results and Discussion}
We compare the LRW index and SRW index with other five similarity
indices, including three local ones: Common Neighbour (CN), Resource
Allocation index (RA) and Local Path index (LP), and two global
ones: Average Commute Time (ACT), Random Walk with Restart (RWR), as
well as the Hierarchical Structure method (HSM). A brief
introduction of each algorithm is shown as follow:

(i) CN: For a node $x$, let $\Gamma(x)$ denote the set of neighbours
of $x$. By common sense, two nodes, $x$ and $y$, are more likely to
have a link if they have more common neighbours. The simplest
measure of this neighbourhood overlap is the directed count, namely
\begin{equation}
s^{CN}_{xy}=|\Gamma(x)\cap \Gamma(y)|.
\end{equation}

(ii) RA: Consider a pair of nodes, $x$ and $y$, which are not
directly connected. The node $x$ can send some resource to $y$, with
their common neighbours playing the role of transmitters. Assuming
that each transmitter has a unit of resource and will equally
distribute it between all its neighbours, the similarity between $x$
and $y$, defined as the amount of resource $y$ received from $x$, is
\cite{Linyuan2009}:
\begin{equation}
s^{RA}_{xy}=\sum_{z\in \Gamma(x)\cap \Gamma(y)}\frac{1}{k_z}.
\end{equation}
Clearly, this measure is symmetric, namely $s_{xy}=s_{yx}$. Note
that this index is equivalent to the two-step LRW, where
\begin{equation}
\pi_{xy}(t=2)=\sum_{z\in
\Gamma(x)\cap\Gamma(y)}\frac{1}{k_x\cdot{k_z}}.
\end{equation}
Former analysis showed that RA preforms best among all the
common-neighbour-based indices \cite{Linyuan2009}.

(iii) LP: This index takes consideration of local paths, with wider
horizon than CN. It is defined as\cite{Linyuan20092}:
\begin{equation}
S^{LP}=A^2+\epsilon A^3,
\end{equation}
where $S$ denotes the similarity matrix, $A$ is the adjacency matrix
and $\epsilon$ is a free parameter. Clearly, this measure
degenerates to CN when $\epsilon=0$. Refs.
\cite{Linyuan2009,Linyuan20092} show that LP, as a semi-local index,
is a good trade-off between effectiveness and efficiency.

(iv) ACT: Denote by $m(x,y)$ the average number of steps required by
a random walker starting from node $x$ to reach node $y$, the
average commute time between $x$ and $y$ is $n(x,y)=m(x,y)+m(y,x)$,
which can be computed in terms of the Pseudoinverse of the Laplacian
matrix $L^{+}$\footnote{$L=D-A$, where $D$ is the degree matrix with
$D_{ij}=\delta_{ij}k_{i}$.}, as \cite{Klein1993}:
\begin{equation}
n(x,y)=E(l_{xx}^{+}+l_{yy}^{+}-2l_{xy}^{+}),
\end{equation}
where $l_{xy}^{+}$ denotes the corresponding entry in $L^{+}$.
Assume that two nodes are considered to be more similar if they have
a smaller average commute time, then the similarity between the
nodes $x$ and $y$ can be defined as the reciprocal of $n(x,y)$,
namely
\begin{equation}
s^{ACT}_{xy}=\frac{1}{l_{xx}^{+}+l_{yy}^{+}-2l_{xy}^{+}}.
\end{equation}

(v) RWR: This index is a direct application of the PageRank
algorithm \cite{Brin1998}. Considering a random walker starting from
node $x$, who will iteratively move to a random neighbour with
probability $c$ and return to node $x$ with probability $1-c$, and
denoting by $q_{xy}$ the probability this walker locates at node $y$
in the steady state, then we have
\begin{equation}\label{rwr}
\vec{q_x}= c P^{T} \vec{q_x}+(1-c)\vec{e_x},
\end{equation}
where $\vec{e_x}$ is an $N\times 1$ vector with the $x^{th}$ element
equal to $1$ and others to $0$. The solution is straightforward, as
\begin{equation}
\vec{q_x} = (1-c) (I-cP^{T})^{-1}\vec{e_x}.
\end{equation}
Accordingly, the RWR index is defined as
\begin{equation}
s^{RWR}_{xy}=q_{xy}+q_{yx}.
\end{equation}

(vi) HSM: The hierarchical structure of a network can be represented
by a dendrogram with $N$ leaves and $N-1$ internal nodes. Each
internal node $r$ is associated with a probability $p_r$ and the
connecting probability of a pair of nodes is equal to $p_m$ where
$m$ is the lowest common ancestor of these two nodes. To predict
missing links with this method we first sample a large number of
dendrograms with probability proportional to their likelihood. And
then calculate the mean connecting probability
$\langle{p_{ij}}\rangle$ by averaging the corresponding probability
$p_{ij}$ over all sampled dendrograms. A higher
$\langle{p_{ij}}\rangle$ indicates a higher probability that nodes
$i$ and $j$ are connected \cite{Clauset2008}.

\begin{figure}
\begin{center}
\includegraphics[width=8cm]{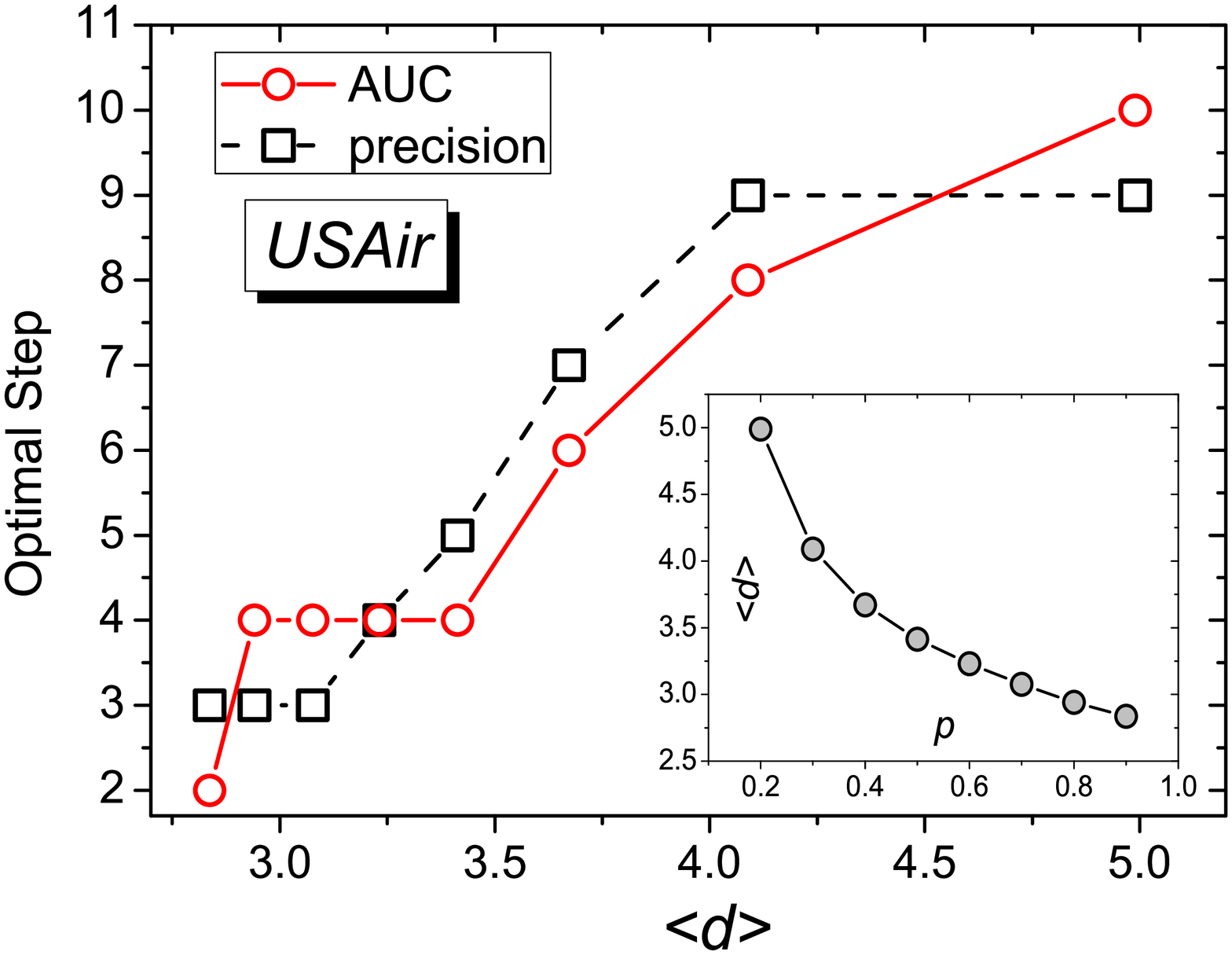}
\includegraphics[width=8cm]{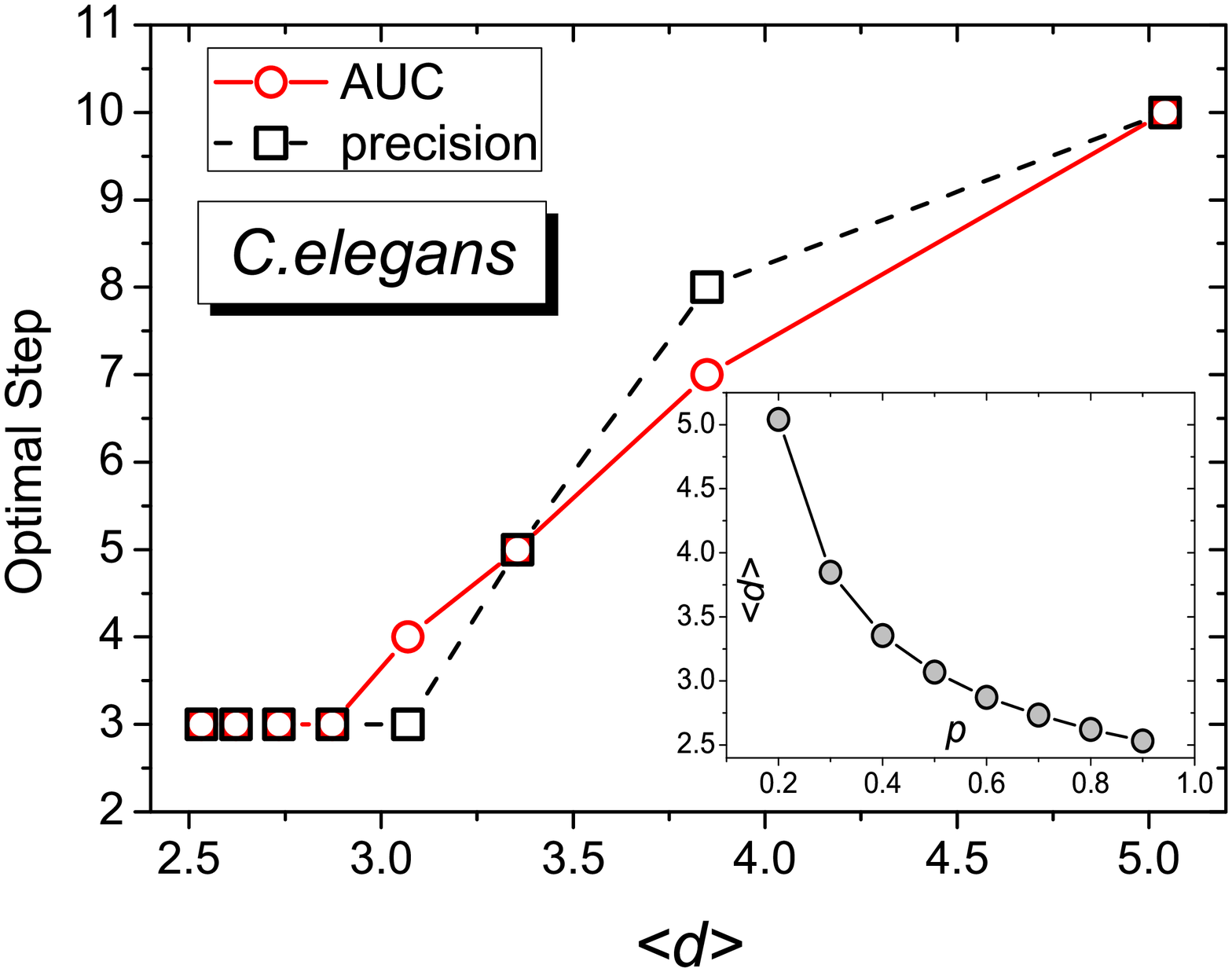}
\caption{(Color online) A positive correlation between the average
shortest distance, $\langle{d}\rangle$, and the optimal step of the
LRW method. The eight points from left to right correspond to the
cases with $p$ from 90\% to 20\%, respectively. The insets show the
dependence of $\langle{d}\rangle$ on the size of the training
set.}\label{optimal}
\end{center}
\end{figure}

The results of these eight methods on five real networks are shown
in Table~\ref{compare}. For each network, the training set contains
90\% of the known links. Generally speaking, the global indices
perform better than the local ones. And our proposed indices, LRW
and SRW, can give overall better predictions than the other methods
for both AUC and precision. Compared with LRW index, the SRW index
can lead to an even higher accuracy. The dependence of accuracy on
the proportion of training set, labeled by $p$, in USAir network and
C.elegans network\footnote{In order to ensure the training set is
connected, the edges should be no less than $N-1$. Therefore, as
shown in Fig.~\ref{train}, not all the investigated networks can be
divided with a 20\%-80\% ratio.} is shown in Fig.~\ref{train}. The
results indicate that the advantage of LRW index and SRW index are
not sensitive to the density of the network.

Interestingly, when predicting with the LRW index, as shown in
Fig.~\ref{optimal}, we find a positive correlation between the
optimal step and the average shortest distance. For example,
$\langle{d}\rangle$ of USAir and C.elegans are very small, no more
than 3, their optimal steps are also small, 2 and 3 respectively in
the case of $p=0.9$. However, in the power grid with
$\langle{d}\rangle\approx{16}$, its AUC keeps increasing at the
beginning and reaches a near optimum at step 16, where one more step
leads to only 0.2\% improvement. We also find that with the
decreasing of $p$, the optimal step increases. This is because the
removal of links to the probe set will increase $\langle{d}\rangle$,
as shown in the insets of Fig.~\ref{optimal}.

Besides high accuracy, the low computation complexity is another
important concern in the design of prediction algorithm. Generally
speaking, the global indices have a higher complexity than the local
indices. As we known, the time complexity in calculating the inverse
or pseudoinverse of an $N\times{N}$ matrix is $O(N^3)$, while the
time complexity of $n$-step LRW (or SRW) is approximately
$O(N\langle k\rangle^n)$. Science in most networks $\langle
k\rangle$ is much smaller than $N$, LRW and SRW run much faster than
ACT and RWR. This advantage is prominent especially in the huge-size
(i.e. large $N$) and sparse (i.e. small $\langle k\rangle$)
networks. For example, LRW for power grid is thousands of times
faster than ACT, even when $n=10$. In HSM, the process to sample a
dendrogram asks for $O(N^2)$ steps of the Markov chain
\cite{Clauset2008}, and in the worse case, it takes exponential time
\cite{Mossel2005}. Each step consumes a certain time to do some
random selections. In addition, to predict the missing links, a
large number of dendrograms are acquired. In this paper, we sample
5000 dendrograms for each implementation. Therefore, the time
complexity of HSM is relatively high. It can handle networks with up
to a few thousand nodes in a reasonable time, while LRW and SRW are
able to handle such networks containing tens of thousands of nodes.
Note that, although ACT, RWR and HSM have a higher time complexity,
they provide much more information beyond link prediction. For
example, the HSM algorithm can be used to uncover the hierarchical
organization of real networks.

\section{Conclusion}
In this Letter, we proposed two similarity indices for link
prediction based on local random walk, the Local Random Walk (LRW)
index and the Superposed Random Walk (SRW) index. We compared our
methods with six well-known methods on five real networks. The
results show that our methods can give remarkably better prediction
than the three local similarity indices. When comparing with the
three global methods, LRW and SRW can give slightly better
prediction with a lower computational complexity.

\acknowledgments We acknowledge V. Batageli and A. Mrvar for the
Pajek Datasets, A. Clauset for providing the code for HSM algorithm,
which can be download from his homepage, and Tao Zhou and Jian-Guo
Liu for their helpful suggestions. This work is partially supported
by the Swiss National Science Foundation (200020-121848), the Future
and Emerging Technologies programmes of the European Commission
FP7-COSI-ICT (LIQUIDPUB project, 213360 and QLectives project,
231200) and the National Natural Science Foundation of China
(60973069).

\end{document}